\documentclass[twocolumn,showpacs,amsmath,amssymb,aps,prd,floats]{revtex4-1}
\usepackage{epsfig}
\usepackage{bm}
\usepackage{graphicx}%
\usepackage{color}
\usepackage{xspace}

\newcommand{\fermi}{\textit{Fermi}-LAT\xspace}

\def\apj{The Astrophysical Journal}
\def\apjl{The Astrophysical Journal Letter}

\def\prd{Physical Review D}

\def\aap{Astronomy and Astrophysics}

\def\mnras{Mon. Not. R. Astron. Soc.}
\def\jcap{JCAP}
\def\apss{Astrophysics and Space Science}

\begin{document}

\title{Indication of a Local ``Fog'' of Sub-Ankle UHECR}
\author{Ruo-Yu Liu$^1$, Andrew M. Taylor$^2$, Xiang-Yu Wang$^3$ and Felix A. Aharonian$^{1,2}$ }
\affiliation{$^1$Max-Planck-Institut f\"ur Kernphysik, Saupfercheckweg 1, 69117 Heidelberg, Germany}
\affiliation{$^2$Dublin Institute for Advanced Studies, 31 Fitzwilliam Place, Dublin 2, Ireland}
\affiliation{$^3$School of Astronomy and Space Science, Nanjing University, Nanjing 210093, China}

\begin{abstract}
During their propagation through intergalactic space, ultrahigh energy cosmic rays (UHECRs) interact with the background radiation fields. These interactions give rise to energetic electron/positron pairs and photons which in turn feed electromagnetic cascades, contributing to the isotropic gamma-ray background (IGRB). The gamma-ray flux level generated in this way highly depends upon the UHECR propagation distance, as well as the evolution of their sources with redshift. Recently, the \fermi collaboration reported that the majority of the total extragalactic gamma-ray flux originates from extragalactic point sources. This posits a stringent upper limit on the IGRB generated via UHECR propagation, and subsequently constrains their abundance in the distant Universe. Focusing on the contribution of UHECR at energies below the ankle within a narrow energy band ($(1-4)\times 10^{18}$~eV), we calculate the diffuse gamma-ray flux generated through UHECR propagation, normalizing the total cosmic ray energy budget in this band to that measured. We find that in order to not over-produce the new IGRB limit, a local ``fog'' of UHECR produced by nearby sources { may} exist, with a possible non-negligible contribution from our Galaxy. Following the assumption that a given fraction of the observed IGRB at 820~GeV originates from UHECR, we obtain a constraint on the maximum distance for the majority of their sources. With other unresolved source populations still contaminating the new IGRB limit, and UHECR above the ankle invariably contributing also to this background, the results presented here are rather conservative.
\end{abstract}

\maketitle

\section{Introduction}
The origin of ultrahigh energy cosmic rays (UHECR, $E>10^{18}$~eV) is a long-standing puzzle. Generally, the community has made a consensus that cosmic rays above  $\sim 4\times 10^{18}$~eV (which is known as the ``ankle'') come from extragalactic sources and cosmic rays below $3\times 10^{15}$~eV (which is known as the ``knee'') stem from our Galaxy, while the origin of cosmic rays in between these two energies are still under debate. 

One indirect way to study the origin of these cosmic rays is to utilize the interactions they undergo with cosmic background radiation during their propagation. Electron/positron pairs and gamma-rays are generated through Bethe-Heitler (BH) pair production and photopion production interactions with the cosmic microwave background(CMB) and extragalactic background light(EBL). These secondary particles proceed to initiate electromagnetic (EM) cascades, resulting in the energies lost by cosmic rays being deposited into lower energy gamma rays, predominantly in the range around 100~MeV to 100~GeV, contributing to the diffuse gamma-ray background\cite{Berezinsky75, Coppi97}. The flux of the latter not only depends on the total amount of cosmic ray energy injected, but also on its spatial distribution. Cosmic rays injected at high redshifts can convert more energy to gamma-rays due to the denser background photons and longer distances available for the cascade to develop. Subsequently, the flux level of the generated gamma-ray background can provide information on the spatial distribution of the UHECR sources.

The Fermi Large Area Telescope (LAT) has measured an angle--averaged spectrum of diffuse gamma-ray emission from high Galactic latitudes in the energy range from 100~MeV to 820~GeV based on 50 months of sky-survey observations. Such emission is referred to as the isotropic gamma-ray background (IGRB, \cite{FermiIGRB15}). This background contains two key components: unresolved extragalactic gamma-ray point sources; and truly diffuse contributions from EM cascades initiated by both UHECRs (mentioned above) and gamma-ray absorption on the EBL. Thus, the measured IGRB flux provides a conservative upper limit for the UHECR fed cascade emission.  

Recently, the \fermi collaboration performed an analysis of the extragalactic gamma-ray data set, finding that extragalactic gamma-ray point sources, including the unresolved ones, contribute at least $86^{+16}_{-14}\%$ of the overall extragalactic gamma-ray background (EGB) consisting of IGRB and resolved discrete sources above 50\,GeV, most of which are blazars \cite{Fermiblazar15}. Since the contribution of unresolved point sources are also counted in the IGRB, their subtraction from the IGRB provides a new upper limit on the cascade flux contribution from UHECR.
According to the source count distribution of extragalactic gamma-ray point sources, obtained in \cite{Fermiblazar15} with their benchmark model, 
we find that unresolved point sources constitute about $46\%$ of the total EGB above 50~GeV. Given that the total flux of the EGB and IGRB around 50~GeV are, respectively, $1.9\times 10^{-7}\rm ~GeV~cm^{-2}~s^{-1}~sr^{-1}$ and $1.1\times 10^{-7}\rm ~GeV~cm^{-2}~s^{-1}~sr^{-1}$ (\cite{FermiIGRB15}, model A), subtracting the contribution by unresolved point sources from the measured IGRB provides an upper limit on the non-point-source IGRB component at the level $2\times 10^{-8}\rm ~GeV~cm^{-2}~s^{-1}~sr^{-1}$ around 50\,GeV. When considering the uncertainty in the obtained source count distribution\cite{Fermiblazar15}, this upper limit is bound by the range $0-4\times \rm 10^{-8}~GeV~cm^{-2}~s^{-1}~sr^{-1}$.

In this paper, we will study the constraint of such an upper limit on the spatial distribution of UHECR sources. The rest of the paper is organized as follows. In Section II, we present our method to calculate UHECR flux and diffusive gamma--ray flux generated by the initiated EM cascade process. We show our results In Section III. In Section IV, we give the discussion and conclusion.

\section{Method}
We calculate the diffuse gamma-ray flux generated via UHECR propagation, on the premise of explaining the observed UHECR flux. Since the fraction of energy converted from cosmic rays to EM cascades also depends on the cosmic ray composition, which above the ankle appears to become heavier \cite{PAO_ICRC15,TA15_composition}, to make our results more conservative we only consider UHECRs in a narrow energy band below the ankle (i.e., $(1-4)\times 10^{18}\,$eV), where the chemical composition is indicated to be proton dominated \cite{PAO_ICRC15,TA15_composition}. Protons in this energy range primarily lose energy via BH interactions. Although only a small fraction ($\sim 2m_e/m_p$) of a proton's energy is transferred to secondary electron/positron pairs in each interaction, the large cross section ($\sim m$barn) ensures that the proton energy loss length at these energies is below the Hubble horizon scale \cite{Berezinsky88, Hooper07}. Subsequently, a significant fraction of proton's energy is passed to secondaries after propagating a sufficiently long  distance, initiating EM cascades and invariably contributing to the IGRB. 

The energy distribution of BH secondaries ($E^2dN/dE$) is broad with a maximum around $10^{-3}E_p$. Provided the cascade has sufficient time to develop, the final spectrum does not keep memory of the details of the distribution of the BH pairs, resulting in the formation of a ``universal spectrum''\cite{Coppi97, Berezinsky75, Berezinsky11}. For the case of more local sources,  when the cascade is not fully developed,
the accurate distribution of BH pairs should be carefully included in the treatment\cite{Stanev00, Kelner08}.

We assume that the injection spectrum of protons above $10^{17}\,$eV takes the form of $f(E)\propto E^{-p}{\rm exp}(-E/E_{\rm max})$, where $E$ is the injected proton energy, $E_{\rm max}=10^{19}\,$eV is the cutoff energy,
and $p$ is the power-law index of the injection spectrum. A range of injection spectra are considered in this work, focusing on fits to data at energies below the ankle. On top of these spectral shape properties, the normalization of the source flux dictates the local energy production rate of UHECRs, $\dot{W}_{p,0}$, whose evolution with redshift is described by, $S(z)$. Provided the UHECR source distribution can be safely approximated as continuous, the total flux of propagated protons from the entire universe can be calculated by
\begin{equation}\label{inte_flux}
\frac{dN}{dE_{\rm ob}}=\frac{1}{4\pi}\int_{z_{\rm min}}^{z_{\rm max}} \frac{\dot{W}_{p,0}}{1+z}S(z)Cf(E)\frac{dE}{dE_{\rm ob}} \frac{cdz}{H(z)}
\end{equation}
where $H(z)$ is the Hubble rate at redshift $z$. Sources within the redshift range between $z_{\rm min}=0.001$, and $z_{\rm max}=5$ are considered. Here, $C$ is the normalization factor of the spectrum, satisfying the relation $\int Cf(E)EdE=1$. The measured proton energy $E_{\rm ob}$ relates to that of the initial proton energy $E$ via a numerical proton energy loss description, which follows the analytic parameterization method proposed in \cite{Kelner08} to treat the interaction between cosmic ray protons and background photons. Along with this, the obtained energy distribution of secondaries produced as a by-product of the proton propagation, enters as source terms in the EM calculations. In both UHECR propagation and cascade development calculations below, we consider CMB and the EBL modeled by \cite{Finke10} as target photons.

To treat cascade processes properly, we utilize a Monte-Carlo simulation to trace the development of cascades. We consider electron/positron pair production through the annihilation of high energy photons with background photons ($\gamma\gamma \to e^+e^-$) and inverse Compton scattering of high energy electrons off background photons ($\gamma e\to \gamma e$). Other processes such as double pair production and triplet production are not important and hence are not taken into account here. We adopt the cross sections and the energy distribution of emitting particles given by \cite{Aharonian83_epp} to treat the annihilation process, while those of inverse Compton scattering are referred to \cite{Blumenthal70}. 

In our calculations a rectilinear treatment of the particle propagation is used. However, we note that the arriving UHECR spectrum is independent of whether the propagation mode is rectilinear or diffusive provided a continuous distribution of sources is assumed\cite{Aloisio04}, with the energy losses simply dictated by the propagation time. A proton of $10^{18}\,$eV ($4\times 10^{18}\,$eV) will lose about $10\%$ ($50\%$), $30\%$ ($70\%$) and $70\%$ ($90\%$) of its energy to EM particles during its propagation to earth if injected at $z=0.5,~1,~3$ respectively\cite{Wang11}. With UHECR passing a significant fraction of their energy flux into cascade gamma-rays, the ratio of the measured energy flux in UHECR at $10^{18}$~eV, relative to that in the IGRB at an energy of tens of GeV, can provide a useful insight. Ignoring reductions to the true IGRB, this ratio is close to unity, with both energy fluxes sitting at a level of $\sim 10^{-7}\,\rm GeV\,cm^{-2}\,s^{-1}\,sr^{-1}$. Severe tension between the true IGRB is, therefore, naturally expected once the IGRB is reduced to $\sim 20\%$ of this level through the removal of the unresolved sources contribution.

\begin{center}
\begin{figure*}[htbp]
\includegraphics[width=0.3\textwidth]{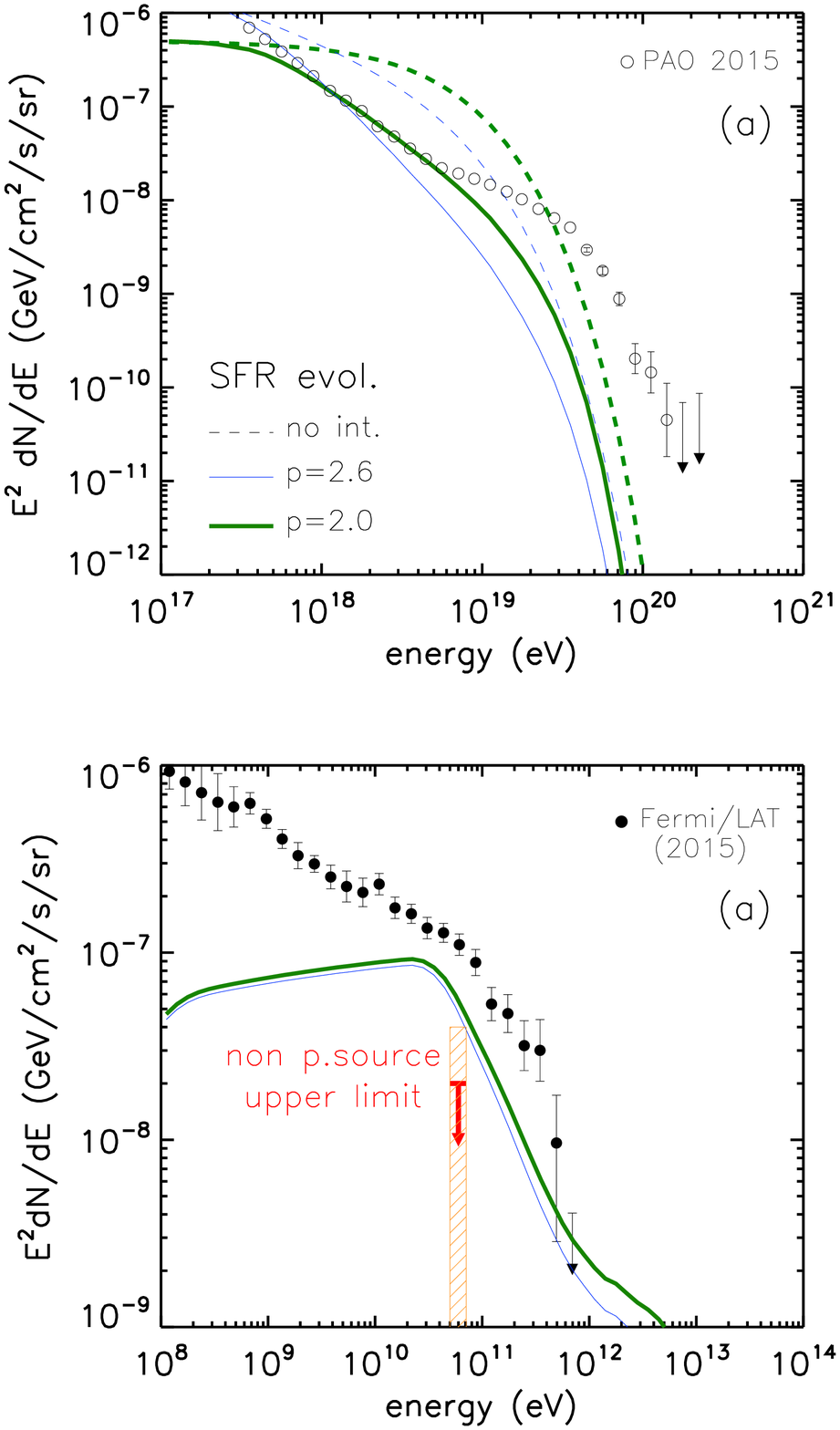}
\includegraphics[width=0.3\textwidth]{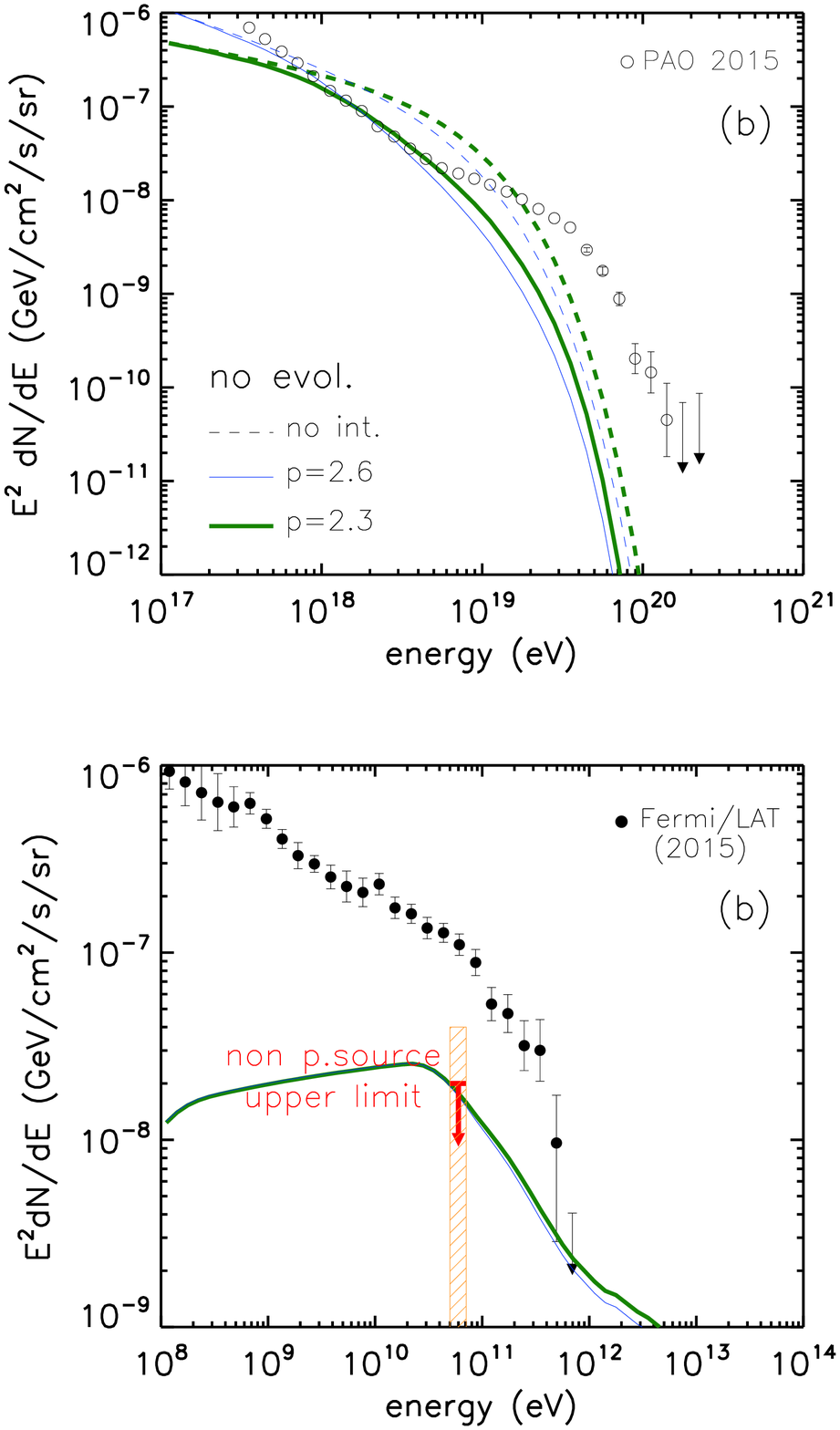}
\includegraphics[width=0.3\textwidth]{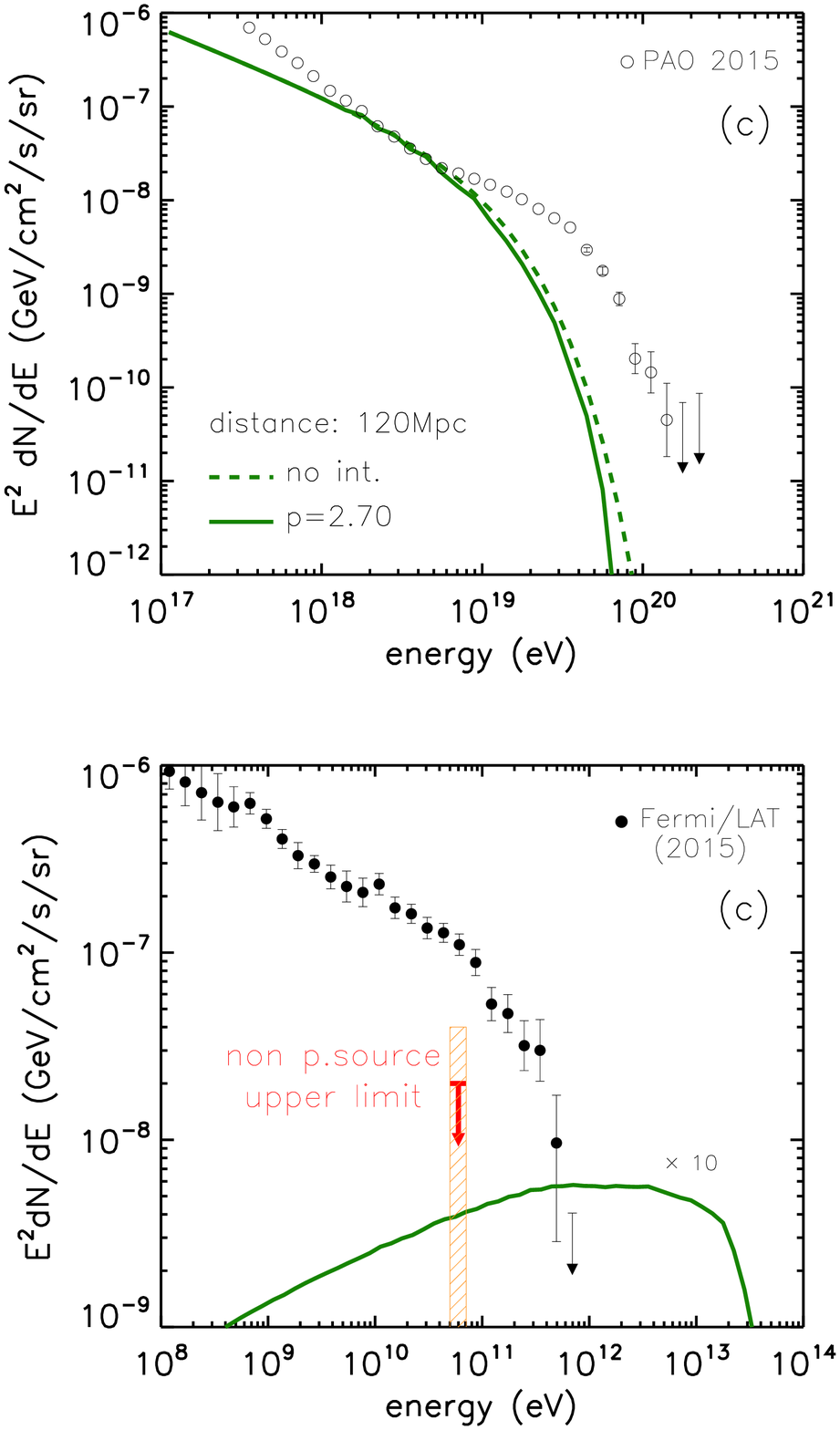}
\caption{Fitting to UHECR spectrum below the ankle and the corresponding diffuse gamma-ray flux initiated by CR propagation with different source distributions ({\bf left}: (a) SFR evolution; {\bf middle}: (b) no evolution; {\bf right}: (c) sources located at 120\,Mpc). In the upper panels, the green solid lines represent the best-fit UHECR fluxes for each source distribution considered, while the dashed lines represent the unattenuated flux. The thin blue lines show the results for a soft injection spectrum of $p=2.6$, normalized to the data at 1\,EeV. Hollow circles show the PAO\cite{PAO_ICRC15} data. The adopted values of the power-law index $p$ and the local energy production rate are provided within the figure. The lower panels show the corresponding diffuse gamma-ray emission resulting from the cascade initiated by UHE protons,  with thick lines and thin lines are respectively for best-fit case and $p=2.6$ case. The black filled circles show the IGRB measured by Fermi/LAT(\cite{FermiIGRB15}). The IGRB upper limit for the non-point-source component (or the truly diffuse component) are shown as a red bar with an arrow. The orange hatched region represents the uncertainty of the limit due to the uncertainties in the obtained source count distribution (i.e., $dN/dS$). The cascade flux in the right panel is multiplied by 10.}
\label{UHECR_IGRB}
\end{figure*}
\end{center}

\section{Result}
The results of our UHECR and cascade calculations are presented in the left and middle panels of Fig.~\ref{UHECR_IGRB}. A range of spectral indices are considered, which are all found to provide a reasonable fit to the data measured by the Pierre Auger Observatory\cite{PAO_ICRC15}. Solid lines represent the arriving UHECR flux while dashed lines represent the unattenuated flux. The calculated cascade fluxes are shown in the lower panels. In panel (a), we assume that the redshift evolution of the UHECR source density follows that of the star formation rate (SFR, \cite{Hopkins06, Yuksel07}). In this case the mean value for the source redshift is $z=1$, with $\sim 40\%$ of the unattenuated flux being lost to EM particles through propagation\footnote{Note that EM particles also suffer adiabatic cooling so not all the energy flux goes into gamma rays.}. The cascade flux is significantly higher than the non-point-source IGRB upper limit, reaching the level of the total measured IGRB. This result is consistent with previous studies\cite[e.g.][]{Ahlers10a, Gelmini12}. In panel (b), results for the case of no evolution in the source density with redshift are shown.  In this case, a larger fraction of UHECRs arrive from lower redshift sources, reducing the energy losses experienced en-route, resulting in less spectral steepening than that for case (a). A range of softer injection spectra are considered for this case than that for SFR evolution. Note that the diffuse gamma--ray flux is not sensitive to the injection spectrum index for the narrow energy band case we consider. This is demonstrated in the figures, with the cascade flux being comparable, regardless of the source index, $p$. Due to the reduced number of sources at high redshift for case (b) relative to case (a), with a mean source redshift of $z=0.6$, only about 20\% of the unattenuated flux is lost to EM particles. The diffuse gamma-ray flux in this case, however, is still marginally above the non-point source IGRB upper limit. { We note that the obtained flux is not strongly dependent on the maximum source redshift, set to $z_{\rm max}=5$ in our calculations, due to the increased source distance and reduced increase in comoving volume at high redshift.}

It is worth highlighting that when calculating the UHECR flux from the entire universe, we scale the energy production rate in the distant universe with the local energy production rate (see Eq.\ref{inte_flux}), as most other authors in the literature have done. The underlying assumption for this treatment is the existence of a uniform and continuous distribution of UHECR sources throughout the whole universe. This may well be established on large spatial and temporal scales. However, it is perhaps unlikely that we are in such an ``average'' place where the local production rate equals the large spatial and temporal scale mean value.

A natural solution preventing UHECR losses over-producing the new IGRB limit is to attribute UHECRs below the ankle to nearby extragalactic objects, or even to our Galaxy{ \cite{Jokipii87, Biermann93b, Budnik08, Ptuskin10}}. In the right panel (c) of Fig.~\ref{UHECR_IGRB}, we consider just such a scenario in which all UHECR sources are located locally, at a distance of 120\,Mpc. Given the proximity of such sources, the fraction of cosmic ray power channeled into gamma-rays is reduced. Along with this, the distance is also insufficient for full cascade development, such that a significant fraction of the energy in the EM population remains in very high energy ($>100\,$GeV) particles, and hence the resulting gamma-ray spectrum peaks at higher energies. The part of the IGRB most constraining the source distance in this scenario comes from the true TeV IGRB.  Fermi-LAT only gives an upper limit of $4\times 10^{-9}\,\rm GeVcm^{-2}s^{-1}sr^{-1}$ for IGRB at 820\,GeV, which should be still composed of both unresolved sources and truly diffuse emission. 
The actual true IGRB upper limit at 820\,GeV highly depends on the spectrum of the unresolved sources which, however, remains unclear.

\begin{center}
\begin{figure}[htbp]
\includegraphics[width=0.4\textwidth]{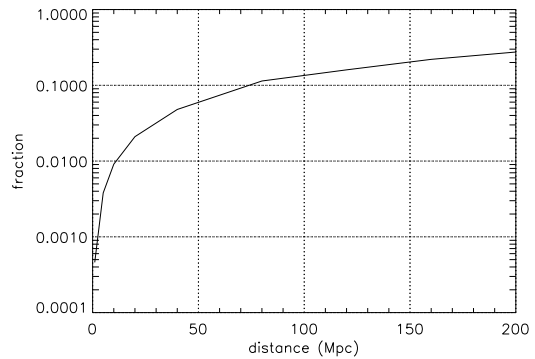}
\caption{The inferred fraction of the measured IGRB upper limit at 820\,GeV caused by $\sim 10^{18}\,$eV cosmic ray propagation as a function of source distance. The energy budget of the cosmic rays is normalized with the measured UHECR spectrum from $10^{18}\,$eV to $4\times 10^{18}$\,eV. { The results are based on the assumption of rectilinear propagation of particles.}}
\label{fraction}
\end{figure}
\end{center}

 { To investigate the effect of a further reduction in such limits, we next explore how the fraction of the upper limit, produced through cascades initiated by UHECRs below the ankle, depends on the source distance. As we can see in Fig.~\ref{fraction}, if the contribution from truly diffuse emission can be constrained down to a level of 10\% of the measured upper limits at 820\,GeV, the main sources of UHECRs below the ankle should be located within 100\,Mpc and if the fraction is constrained down to the 1\% level, the sources should be closer than 10\,Mpc. Finally, a Galactic proton EeVatron may be required if the unresolved sources constitute more than 99.9\% of the IGRB at 820\,GeV. Future observation of Fermi-LAT and the next generation very-high-energy gamma-ray detectors such as CTA, which has much higher sensitivity, will provide us a deeper insight into this issue by resolving a larger number of point sources from the IGRB at the TeV scale. We note that the calculation here is based on the assumption of no magnetic field, with the rectilinear propagation of particles. For this fixed source distance scenario, considering the intergalactic magnetic field would effect the spectral results. The obtained flux could be increased or decreased, depending on the specific configuration of the magnetic field. However, a detailed study is beyond the scope of this work.}

In scenario (c), distant UHECR sources could still also be present, but their energy production rate should be smaller than that from a simple scaling with the local energy production rate. In other words, this scenario encapsulates the consideration of an over-density of UHECR sources in the local universe. In such a situation, we are embedded in a local ``fog'' of cosmic rays produced by nearby sources, which exceeds the true mean (sea) value. We note that a negative evolution of UHECR sources was suggested in \cite{Taylor15}, which may also resolve the contradiction. However, most of UHECR candidate sources show a  positive and strong evolution \cite[e.g.][]{Wall05, Guetta07, Yuksel08}, although a few kinds of extragalactic objects, such as low-luminosity high-synchrotron peaked BL Lacs, do seem to possess a negative evolution\cite{Ajello14}. 


\section{Discussion and Conclusion}
In this paper, we have studied the constraint of the non-point-source component of the IGRB on the origin of UHECRs below the ankle.  We obtained an upper limit for the non-point-source component of the IGRB. We then fitted the UHECR spectrum within a narrow energy band between $10^{18}\,$eV and $4\times 10^{18}\,$eV, calculating the accompanying diffuse gamma-ray emissions resulting from the produced UHE secondaries. We found that scaling the energy production rate of cosmic rays in the distant universe with the local energy production rate, and adopting a reasonable redshift evolution function, the diffuse gamma-ray flux generated through cascades overshoots the upper limit of the non-point-source IGRB. To resolve this conflict, we propose that an over-density exists in the energy production rate of UHECRs in the nearby universe, with these sources dominating the observed UHECR flux at $\sim 10^{18}$eV. In this case, the short propagation times prevents a significant fraction of the CR energy flux being converted into low energy gamma-rays, hence reducing the diffuse gamma ray flux contribution from UHECR fed cascades. Furthermore, a high cosmogenic neutrino flux is not expected in this scenario. The distance of these cosmic ray sources can be further constrained if we have a better understanding on the non-point-source component of the IGRB around 1~TeV.  

We note that our results are quite conservative since the non-point-source IGRB upper limit will be contributed by other processes. First, the UHECRs above the ankle, which are intentionally neglected in our calculations (to be maximally model-independent and robust), also initiate cascades through pair production and photopion production interactions, and would be expected to further contribute to the IGRB. Second, cosmic rays below $10^{18}\,$eV with an extragalactic origin will also contribute to the diffuse gamma-ray background through cascades, especially for those injected at high redshift. Indications of such a component come from the KASCADE experiment, which revealed a light component of cosmic rays with  a spectral index of $2.8$ between $10^{17}-10^{18}\,$eV\cite{KASCADE13}. Although we considered cosmic ray injection down to $10^{17}$eV in our calculations, the obtained total flux in this energy range is below the level of this component. 

On top of further UHECR contribution to the diffuse gamma-ray background, other point sources must also contribute, most notably misaligned AGN whose contribution to the new IGRB derived is expected to be considerable { \cite{Inoue11b, DiMauro14, Hooper16}}. Moreover, the very high energy photons from those whose intrinsic spectra extend beyond $100\,$GeV are absorbed by the EBL, with this energy reprocessed into lower energy gamma-rays via cascades\cite{Coppi97}. Considering these components will further aggravate the tension between the Fermi's data and the theoretical diffuse gamma-ray flux induced by UHECR propagation. 


{ On the other hand, the tension between the Fermi-LAT data and the theoretical flux could potentially be overestimated in certain cases. For example, if UHECRs mainly come from the resolved gamma-ray point sources and the intergalactic magnetic field is sufficiently weak so as not to deflect UHECRs outside the jet cone, the subsequent electron/positron pairs can be produced within the lines of sight of their sources. The secondary gamma-ray photons generated in cascades from these pairs will then constitute part of the emissions from these point sources and hence the obtained non-point-source IGRB upper limit may be smaller than the true value, alleviating the tension. However, one should heed caution with this scenario, since the majority of resolved point sources that contribute EGB are blazars which present strong variabilities, while the lightcurve of UHECR-fed cascades should be smooth on short ($<$yr) time-scales  \cite{Prosekin12}.}

{ Given the relatively large uncertainty of the current measurement, the present tension between the theoretical flux of UHECR-fed cascades and the non-point-source IGRB is at the $\sim 1\sigma$ confidence level. Future observations with higher statistics will provide a deeper insight in this issue. If it can be confirmed that $\sim 90\%$ of the IGRB is contributed by point sources, the need for the domination of local CR sources is unavoidable.}

It is lastly interesting to highlight that if UHECRs below the ankle are mainly from nearby sources, isotropization of cosmic rays by Galactic magnetic field and intergalactic magnetic field \cite{Kumar14, Takami16} is required to avoid considerable anisotropy \cite{PAO13_ani}. Since intergalactic magnetic fields are potentially insufficient in strength to achieve this \cite{Ryu08, Dolag05, Donnert09}, a large Galactic halo of $\sim 100\,$kpc and $\mu \,G$ magnetic field may instead be required.  
Some recent studies on the hot gases in Galactic halo might support this scenario\cite{Gupta12, Fang13, Miller15}. However, a detailed study on the isotropization in local magnetic fields is beyond the scope of this work. 

\bibliographystyle{apsrev4-1}
\bibliography{ms_revise_PRD.bib}

\end{document}